\documentclass[a4paper,oneside,final,notitlepage,onecolumn]{article}
\usepackage{amsfonts}
\usepackage{epsf}
\usepackage[dvips]{color}
\usepackage[latin2]{inputenc}

\begin{document}

\title{On rigidity of spacetimes with a compact Cauchy horizon}

\author{Istv\'{a}n R\'{a}cz
\\ 
 MTA KFKI, Részecske- és Magfizikai Kutatóintézet,
              Elméleti Főosztály\\
 H-1121 Budapest, Konkoly Thege Miklós út 29-33. Hungary\\
E-mail: istvan@sunserv.kfki.hu }


\maketitle

There is a great variety of solutions of Einstein's equation which
contain a  smooth compact Cauchy horizon. However, all of these
spacetimes  can be considered as being generalizations of the Taub-NUT
solution. In these spacetimes, the Cauchy horizon separates the
globally hyperbolic region from a region which contains closed
timelike curves. It is a common feature of all of these spacetimes
that they admit a `horizon compatible' Killing vector field which is
spacelike in the globally hyperbolic region, null on the  horizon
while the Killing orbits are closed timelike curves in the chronology
violating region.

Since in all particular examples there exists a
Killing vector field it is reasonable to ask:  To what extent the
field equations can guarantee the existence of  such a Killing vector
field? Clearly, this question has a direct relation to the
validity of the strong cosmic  censor hypothesis. This hypothesis of
Penrose claims that all the `physically reasonable' generic
spacetimes are globally  hyperbolic, i.e. they are inextendible Cauchy
developments of `maximal' initial data specifications. In this
respect, if one could show that the existence of a compact Cauchy
horizon is always associated with the presence of a symmetry one would
have an indirect argument supporting the validity of Penrose's cosmic
censor hypothesis.

Actually, a fundamental result of this type was given by Moncrief
and Isenberg  \cite{im1,im2}. They considered analytic  electrovac
spacetimes  possessing a compact Cauchy horizon generated by closed
null geodesics. As shown by the authors there must exist then a
Killing vector field which is tangential to the null geodesic
generators of the horizon and spacelike on the Cauchy development
side. This result of Moncrief and Isenberg had  been established by the
mid
of 80's. It has remained, however, an interesting open problem whether
the analyticity assumption can be replaced by a less restrictive
differentiability condition. It is worth emphasizing that, since the
use of the analyticity assumption is incompatible with the concept of
causality e.g.  in the framework of initial value problem, this issue
has not only of pure mathematical interest.

To construct a `candidate' Killing vector field in the smooth setting
the relation
\begin{equation}
\nabla ^e\nabla _eK^a+{R^a}_dK^d=0  \label{Laplrk}
\end{equation}
can be used, which is known to be satisfied by any Killing vector
field. This is a wave equation for $K^a$ suggesting that our approach
should be based on an initial value problem. Since the horizon is
null, the use of the null characteristic initial value problem seems
to be appropriate where the initial data are specified on a pair  of
null hypersurfaces which intersect on a smooth spacelike 2-surface.

This means that the Cauchy horizon, by itself, does not
comprise a suitable initial data surface for (\ref{Laplrk}). In fact,
the main difficulty one has to face in generalizing the
Isenberg-Moncrief theorem to the smooth setting is that suitably
detailed information about the spacetime metric and the
electromagnetic field is known only on the horizon. To be able to use
the characteristic initial value problem an additional
null hypersurface, on which (at least) the Lie derivative of the
spacetime metric vanishes, was needed to be found.

By making use of a straightforward generalization of the techniques
applied in  Refs.\cite{rw1,rw2}, such a null hypersurface can be
constructed whenever the original Cauchy horizon is
non-degenerate\cite{frw}. In addition, the Lie derivatives of the
spacetime metric and the electromagnetic field with respect to a
suitably chosen vector field, $k^a$, vanish throughout the resulting
bifurcate horizon. This horizon provides an appropriate
initial data surface.  Upon having this surface, it remains to show
that the unique solution $K^a$ of (\ref{Laplrk}), corresponding to
initial data compatible with $k^a$,  will be a Killing vector field in
the associated Cauchy development. It follows, however, that, whenever
$K^a$ fulfills  (\ref{Laplrk}) but is otherwise an arbitrary vector
field, the Lie derivatives of the spacetime geometry and the
electromagnetic field variables satisfy a coupled system of linear and
homogeneous  wave equations in these basic
variables\cite{frw,r}. These type of  equations are known to possess
unique solutions, hence, they possess the identically zero solution
for vanishing initial data.

These results were also found to work in the cases of Klein-Gordon,
Higgs, Yang-Mills, Yang-Mills-dilaton and Yang-Mills-Higgs fields in
Einstein's theory of gravity\cite{r1,r}.\footnote{%
It is worth
emphasizing that in the analytic setting the underlying techniques
provide a generalization not only of the result of Moncrief and
Isenberg but that of the Hawking's black hole rigidity   theorem, as
well.} The following statement, that applies to all of these general
Einstein-matter systems, sums up the argument outlined above:

\medskip
\parindent 0pt

{\bf Theorem:} {\sl  Consider a smooth spacetime with a compact Cauchy
horizon generated by closed null geodesics. Suppose that the horizon
is non-degenerate.  Then  there exists a smooth Killing vector field
$k^a$ in a sufficiently small neighbourhood of the horizon in the
Cauchy development region. The horizon is a Killing horizon with
respect to $k^a$, while this Killing vector field is spacelike off the
horizon.  The matter fields are also invariant.}

\medskip
\parindent 20pt

According to this result the presence of a compact non-degenerate Cauchy
horizon, ruled by  closed null geodesics, is really an artifact of a
spacetime symmetry. This, in turn, supports the validity of
the cosmic censor hypothesis  by demonstrating the
non-genericness of spacetimes possessing such a compact Cauchy
horizon.

\section*{Acknowledgments} Partially supported by OTKA grant T030374.

\end{document}